\documentclass[journal,twocolumn]{IEEEtran}
\usepackage{epsfig,makeidx,color}
\usepackage{amsmath,amssymb,amsthm,bbm}
\usepackage{cite,graphicx}
\usepackage{enumerate}
\usepackage{hyperref}
\hypersetup{
	colorlinks = true,
	citecolor=blue,
}

\def\cK{{\cal K}}
\def\cX{{\cal X}}

\def\rT{{\rm T}}

\def\uR{{\mathbb R}}

\def\uE{{\mathbb E}}

\DeclareMathOperator*{\argmin}{\arg\!\min}
\DeclareMathOperator*{\argmax}{\arg\!\max}

\def\be{ \begin{equation} }
\def\ee{ \end{equation} }
\def\bea{ \begin{eqnarray} }
\def\eea{ \end{eqnarray} }
\def\bx{{\bf x}}
\def\by{{\bf y}}

\def\bb{{\bf b}}

\def\bg{{\bf g}}

\def\bs{{\bf s}}
\def\ba{{\bf a}}

\def\bu{{\bf u}}

\def\bv{{\bf v}}
\def\bw{{\bf w}}

\def\bG{{\bf G}}

\def\bR{{\bf R}}

\def\b0{{\bf 0}}

\def\bPsi{{\bf \Psi}}

\def\cC{{\cal C}}

\def\cK{{\cal K}}

\def\cN{{\cal N}}

\def\sH{{\sf H}}

\ifCLASSOPTIONonecolumn
  \newcommand{\figwidth}{0.60\columnwidth}
\else
  \newcommand{\figwidth}{0.93\columnwidth}
\fi

\begin{document}

\title{Data-aided Sensing where Communication 
and Sensing Meet: An Introduction}

\author{Jinho Choi\\
School of Information Technology \\
Deakin University, Australia \\
(e-mail: jinho.choi@deakin.edu.au).}

\date{today}
\maketitle

\begin{abstract}
Since there are a number of Internet-of-Things
(IoT) applications that need to collect data sets
from a large number of sensors or devices
in real-time, sensing and communication need to be integrated
for efficient uploading from devices.
In this paper, we introduce the notion of
data-aided sensing (DAS) where 
a base station (BS) utilizes a subset of data that is already uploaded 
and available to select the next device for 
efficient data collection or sensing.
Thus, using
DAS, certain tasks 
in IoT applications, including federated learning,
can be completed by uploading from
a small number of selected devices.
Two different types of DAS are considered: one is centralized DAS
and the other is distributed DAS.
In centralized DAS, the BS decides the uploading order,
while each device can decide when 
to upload its own local data set among multiple uploading rounds
in distributed DAS. In distributed DAS,
random access is employed where the access
probability of each device is 
decided according to its local measurement for efficient uploading.
\end{abstract}

{\IEEEkeywords
Data-aided sensing;
Internet-of-Things (IoT); Cross-Layer; Federated Learning}

\ifCLASSOPTIONonecolumn
\baselineskip 26pt
\fi

\section{Introduction} \label{S:Intro}

It is expected that
the Internet of Things (IoT) will play 
an important role in a number of applications
in Industry 4.0, including smart cities and factories, in the future
\cite{Gubbi13} \cite{Kim16}.
There could be a number of IoT systems
and it is desirable to build them 
as layered systems with interoperability,
where the bottom layer is responsible for collecting and processing
information or data from devices or sensors
\cite{Fuqaha15}.

For IoT connectivity, 
a number of solutions are studied including
WiFi, cellular IoT, and so on. Compared to others,
cellular IoT can support IoT applications
over a large area. For example, as discussed in \cite{Mang16},
narrowband IoT (NB-IoT) \cite{3GPP_NBIoT} can be used
to support IoT applications over a large area through cellular systems.
In cellular IoT (for the bottom layer
in IoT systems), each base station (BS)
can be used as a data collector from devices
or sensors deployed over a cell.
Since long-term evolution (LTE)
BSs are well deployed, cellular IoT 
might play a crucial role as IoT infrastructure in collecting
a large amount of data from devices including mobile phones 
and sensors.

In a number of IoT applications,
collecting data sets from devices deployed in an area
requires devices' sensing to acquire local measurements or data 
and uploading to a server.
In cellular IoT, devices
are to send their local measurements to a BS.
While sensing and uploading can be considered separately,
they can also be integrated, which leads to
data-aided sensing (DAS) \cite{Choi19}.
In general, DAS can be seen as iterative data collection scheme
where a BS is to collect data sets from devices or 
nodes\footnote{Throughout
the paper, it is assumed
that nodes, devices, and sensors are interchangeable.}
through multiple rounds.
In DAS, the BS chooses a set of nodes in each round
based on the data sets that are already available at the BS
from previous rounds for efficient data collection.
As a result, 
the BS (actually a server that is connected to the BS)
is able to efficiently provide an answer to a given query 
with a small number of measurements compared to random polling.
%(i.e., all nodes' measurements)

In this paper, we introduce the key idea 
and approaches of DAS \cite{Choi19} \cite{Choi19_IoT} \cite{Choi19_WCL}.
In particular, two different DAS schemes, namely
centralized DAS and distributed DAS, are discussed.
Depending on applications,
in DAS,
different objective functions
can be used in selecting the devices
in each round.
We consider an entropy-based objective function
to illustrate the idea of DAS and also discuss
another objective function when measurements
have a sparse representation \cite{Choi19},
where the notion of compressive sensing (CS)
\cite{Donoho06} \cite{Candes06} is exploited.

While the BS decides uploading order in centralized DAS,
each user can decide whether
or not to upload its local data in each round in distributed DAS.
Due to uncoordinated transmissions in distributed DAS,
random access 
is employed with parallel multiple channels,
where the access (or uploading) probability
of each device can be decided by its measurement.
This approach can be applied to federated learning
\cite{FO16} \cite{Yang19} \cite{Park19},
as studied in \cite{Choi19_WCL}.

It is noteworthy that this paper is to introduce
DAS that has been recently studied in \cite{Choi19}
\cite{Choi19_IoT} \cite{Choi19_WCL}.
As a result, this paper can be seen as a review paper,
while it is new to classify DAS into centralized and
distributed ones and introduce them.

{\it Notation}:
Matrices and vectors are denoted by upper- and lower-case
boldface letters, respectively.
The superscript $\rT$
denotes the transpose.
For a set, $\cX$, its complement set is denoted by $\cX^c$.
$\uE[\cdot]$
and ${\rm Var}(\cdot)$
denote the statistical expectation and variance, respectively.
In addition, ${\rm Cov}(\bx)$ represents
the covariance matrix of random vector $\bx$.
$\cN(\ba, \bR)$ and 
$\cC\cN(\ba, \bR)$
represent the distributions of
real-valued Gaussian 
and circularly symmetric complex Gaussian (CSCG)
random vectors with mean vector $\ba$ and
covariance matrix $\bR$, respectively.

\section{System Model}

\begin{figure}[thb]
\begin{center}
\includegraphics[width=\figwidth]{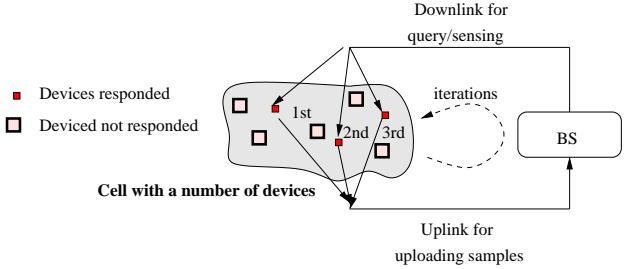}
\end{center}
\caption{Devices deployed in a certain area to collect
local data sets and upload to a base station.}
        \label{Fig:e_das}
\end{figure}

Suppose that there are 
a number of devices (say $K$ devices) within a cell
and each device has a local measurement or data to upload to
a BS that is to collect a data set of
measurements. For example,
each device can be seen as a sensor node that
is deployed over a certain area to collect local environmental data,
or a mobile device  that is associated
with a certain application that needs to learn
from a set of data from a number of mobile devices
(e.g., federated learning \cite{FO16} \cite{Yang19} \cite{Park19}).
When the bandwidth is limited 
with a large $K$, it is difficult to
simultaneously upload all the measurements from devices. 
Thus, there should be
multiple transmissions or uploading rounds to be carried
out as illustrated in Fig.~\ref{Fig:e_das}.

Suppose that there are $L$ parallel multiple 
access channels for uplink transmissions. To upload
all the measurements from $K$ devices,
there might be $\lceil \frac{K}{L} \rceil$ rounds.
There can be a pre-determined transmission
order for sequential polling,
where uploading is carried out regardless
of local data sets.
For efficient uploading,
however, DAS can be employed as in
\cite{Choi19} \cite{Choi19_IoT}
\cite{Choi19_WCL}.
In general, there are two types of DAS approaches.
One is 
\emph{centralized} DAS and the other is 
\emph{distributed} DAS.
In centralized DAS, in each round, the BS decides
the devices that can upload their local 
measurements based on the accumulated measurements
from the previous rounds.
On the other hand, in distributed DAS, each device decides
whether or not to upload its transmission for a given round.
In this paper, we review those DAS approaches.

\section{Principles of DAS}

In this section, we present the notion of DAS
for a specific problem where the BS is to collect 
local measurements.

Suppose that $L = 1$ (i.e., in each round, only one
device can upload its measurement).
Let $x_k$ denote the local measurement of device $k$.
In addition,
denote by $\cX (t)$ the set of local measurements
that are available at the BS after iteration $t$. 
Then, for a certain objective function, $\Theta(\cdot)$,
the BS can decide the next device to upload its local information
as follows:
\be
k (t+1) = \argmin_{k \in \cK^c (t)} \Theta (x_k, \cX(t)),
	\label{EQ:key}
\ee
where $k(t)$ is the index of the device that is to upload
its local measurement at round $t$
and $\cK(t)$ represents the index set of the devices associated with
$\cX(t)$. Clearly, it can be shown that
$$
\cK(t) = \{k(1), \ldots, k (t)\}.
$$
In \eqref{EQ:key}, we can see that
the selected device, say $k(t+1)$,
in each round depends on
the accumulated measurements
from the previous rounds, $\cX(t)$, which shows the key idea of DAS.

The objective function in
\eqref{EQ:key} varies depending on the application.
As an example, let us consider an entropy-based
DAS where 
the BS is to choose the uploading order based on the entropy
of measurements.
Since 
$\cX(t)$ is available at the BS after iteration
$t$, the entropy or information of the remained measurements becomes
$$
\sH(\cX^c (t) \,|\, \cX(t)),
$$
where $\sH(X|Y)$ represents the conditional entropy of $X$
for given $Y$.
Thus,
the objective function can be given as
\be
\Theta(x_k, \cX(t))
=
\underbrace{\sH(\cX^c (t) \,|\, \cX(t))}_{\rm Remained \ Information}
- \underbrace{\sH(x_k \,|\, \cX(t))}_{\rm Updated\ Information},
	\label{EQ:Ent_O}
\ee
which is the entropy gap,
where $\sH(x_k \,|\, \cX(t))$ is the amount of information 
by uploading the measurement from device $k$.
Clearly, for fast data collection 
(or data collection with a small number of devices),
we want to choose device $k$ to minimize the objective function.

Note that
$\sH(\cX^c (t) \,|\, \cX(t))$ is independent of $k$.
Thus, the next device is to be chosen according
to the maximization of conditional entropy is given by
\be
k (t+1) = \argmax_{k \in \cK^c (t)} \sH(x_k\,|\, \cX(t)).
\ee
That is, the next device should have the maximum
amount of information
(in terms of the entropy)
given that $\cX(t)$ is already available
at the BS.

\begin{figure}[thb]
\begin{center}
\includegraphics[width=7cm,height=7cm]{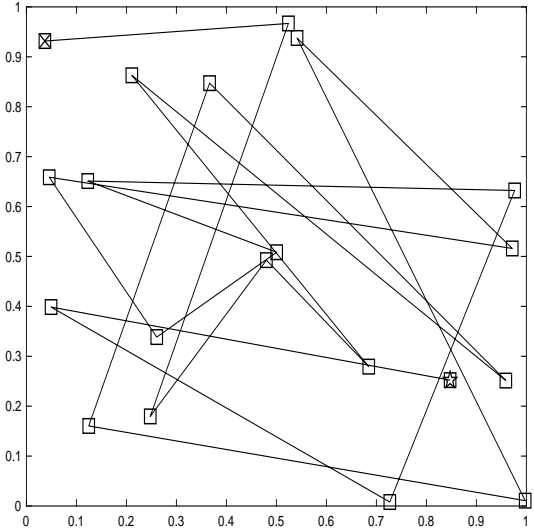} 
\end{center}
\caption{Data Collection based on DAS from 20 devices,
where the first and last device are represented by star and cross markers,
respectively.}
        \label{Fig:1}
\end{figure}

For an illustration,
suppose that there are $K = 20$ devices or nodes,
which are deployed in a unit square,
with their local measurements. 
Assume
that the correlation of measurements is proportional to
the distance between devices, i.e., the correlation coefficient is
$\rho_{i,k} = e^{- ||\bu_i - \bu_k||}$,
where $\bu_k$ denotes the coordination of device $k$.
In Fig.~\ref{Fig:1}, the locations of 20 
devices are represented by square markers,
while the first device, $k(1)$, 
and the last device, $k (K)$, are represented
by star and cross markers, respectively. In addition, each edge 
represents the link between nodes $k(t+1)$ and $k(t)$.
In general, there is no edge 
between two adjacent devices due to a high correlation,
and the next device tends to be far away from the current device
to minimize the entropy gap (as a far device likely has
a highly uncorrelated measurement with the current device's one).

In \cite{Choi19_IoT},
the mean squared error (MSE) 
between the total measurement,
$\bx = [x_1 \ \ldots \ x_K]^\rT$,
and its estimate $\hat \bx_t (x_k, \cX(t-1))$,
which is an estimate of $\bx$ from $x_k$ and $\cX(t-1)$,
is to be minimized in choosing the device 
in round $t$,
i.e.,
\be
\Theta (x_k, \cX(t))
= \uE \left[||\bx - \hat \bx_t (x_k, \cX(t-1)) ||^2 \, |\, 
\cX(t-1) \right],
\ee
where the expectation is carried out over
$\cX^c(t-1)$, which is the subset of measurements
that are not available before round $t$.
Since the measurements are assumed to be Gaussian 
in \cite{Choi19_IoT},
second order statistics can be used for DAS, 
which is called Gaussian DAS.

%\cite{Choi19_WCL}.

\section{DAS for Sparse Signal Sources}

In this section, we consider the
case that measurements have a 
sparse representation and briefly present the approach
in \cite{Choi19}.

Suppose that
the signal vector obtained
from all the nodes' measurements, i.e., $\bx$,
has a sparse representation
\cite{Frag13} \cite{Karakus13} \cite{Wu18}.
Let
$$
[\bx]_k = x_k \in \uR, \ k \in \cK = \{1,\ldots, K\}.
$$
Furthermore, $\bx$ can be represented by
a sparse vector $\bs \in \uR^{M \times 1}$.
To this end, it is assumed that
\be
x_k = \bb_k^\rT \bs,  \ k \in \cK,
        \label{EQ:vbs}
\ee
where $\bb_k \in \uR^{M \times 1}$ is the measurement
vector at node $k$, which is known at the BS.

Denote by $\cK_0$ the set of the
devices uploading their measurements.
To be precise, let
\be
\cK_0 = \{k_1, \ldots, k_N\}, \ k_n \in \cK,
\ee
where $k_n$ represents the index of the $n$th node
uploading local measurement.
The subvector of $\bx$ associated with $\cK_0$ is denoted by
$\bw$ and let $\bPsi = [\bb_{k_1} \ \cdots \ \bb_{k_N}]^\rT$.
Then, it can be shown that
\be
\bw = \bPsi \bs \in \uR^{N \times 1},
        \label{EQ:wPs}
\ee
which can be seen as a random sensing or sampling of $\bx$.
It is known that if
$N \ge C S \log \left( \frac{M}{S} \right)$,
where $C$ is a constant, $\bs$ can be recovered
from $\bw$ under certain conditions of $\bPsi$ in \eqref{EQ:wPs}
\cite{Donoho06} \cite{Candes06} \cite{Eldar12}.
Once $\bs$ is available,
$\bx$ can be obtained from \eqref{EQ:vbs}.
In other words, without collecting all the measurements
from $K$ nodes, it is possible to estimate
$\bx$ from $N$ nodes' measurements.

\begin{figure}[thb]
\begin{center}
\includegraphics[width=\figwidth]{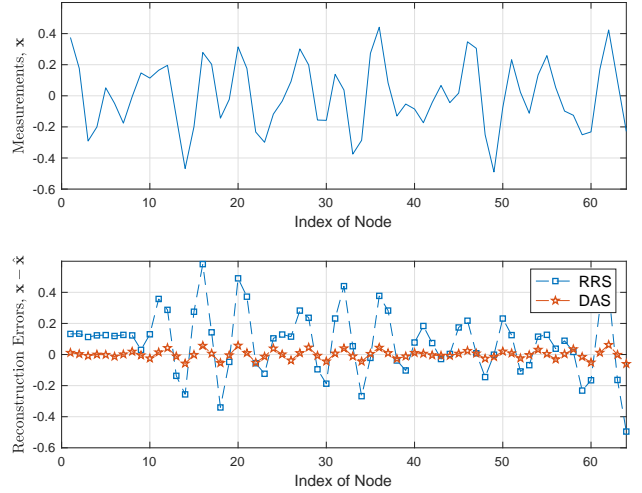}
\end{center}
\caption{The measurements at $K$ nodes (the upper plot)
and the reconstruction errors after $4$ rounds
(the lower plot)
when $K = 64$, $M = 25$, $S = 3$, and $N = 5$.}
        \label{Fig:plt_wav}
\end{figure}

Fig.~\ref{Fig:plt_wav} shows
reconstruction errors at the BS with DAS
and repeated random sensing (RRS) (i.e., random polling)
after 4 rounds together with
the target signal, $\bx$,
when $K = 64$, $M = 25$, $S = 3$, and $N = 5$.
As shown in Fig.~\ref{Fig:plt_wav},
DAS can provide a good estimate
of $\bx$ at the BS with $4 N = 20$ measurements,
while RRS cannot provide a reasonable estimate.

Let $\bv (t)$ and $\bs(t)$ 
be the estimates of $\bx$ and $\bs$ at round $t$
(with data set $\cX(t)$), respectively.
Note that once $\bs(t)$ is obtained,
we have
\be
\bv (t) = [\bb_1 \ \cdots \ \bb_K]^\rT \bs(t).
\ee
In DAS, for given $\cK(t)$,
the node with the most significant
measurement value (in terms of the amplitude)
in round $t+1$ can be chosen as follows:
\begin{align*}
k(t+1) & = \argmax_{k \in \cK^c (t)} |[\bv (t) ]_k|^2 \cr
& = \argmax_{k \in \cK^c (t)} |[\bb_k^\rT \bs(t)|^2.
\end{align*}
Note that $x_k$ and $x_i$ are highly correlated
with each other if the correlation
of $\bb_k$ and $\bb_i$ is high.
This can be taken into account and the
resulting node section criterion for DAS
becomes
\be
k(t+1) = \argmax_{k \in \cK(t)^c}
\min_{i \in \cK(t)}
\frac{ |\bb_k^\rT \hat \bs(t) |^2}{|\bb_k^\rT \bb_i|^2}. 
        \label{EQ:ks3}
\ee

\begin{figure}[thb]
\begin{center}
\includegraphics[width=\figwidth]{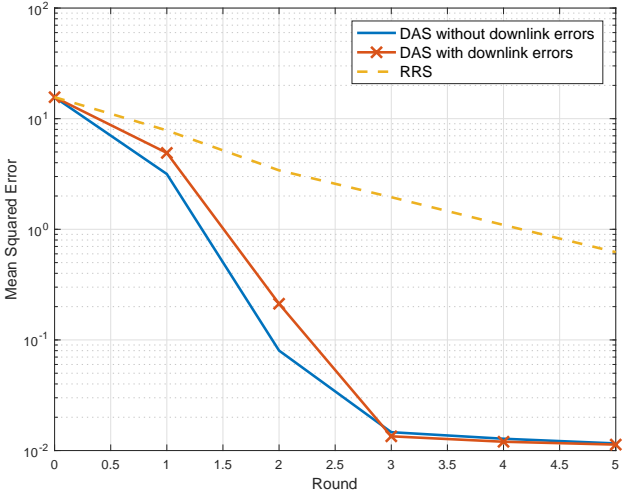}
\end{center}
\caption{Estimation Errors versus rounds
with $K = 300$, $S = 10$, and $L = 10$ parallel multiple access channels
per round.}
        \label{Fig:aplt1}
\end{figure}

While only one node is chosen per round in 
\eqref{EQ:ks3}, we can choose up to $L$ nodes
if there are $L$ parallel multiple access channels.
In Fig.~\ref{Fig:aplt1},
the MSE of the estimate of $\bx$ is shown
as a function of round $t$
with $K = 300$, $S = 10$, and $L = 10$ parallel multiple access channels
per round. Clearly, for a reasonable estimate of $\bx$,
it is shown that DAS needs a smaller number of rounds 
than RRS.
It is noteworthy that the BS
needs to send request signals to the selected
nodes in each round through
downlink channels and there might be 
errors in downlink transmissions (which results in 
uploading from unselected nodes).
As shown in 
Fig.~\ref{Fig:aplt1},
the performance degradation due to downlink
errors is not significant.

\section{Distributed DAS with Random Access}

In centralized DAS,
the BS is to collect measurements
from devices through iterations or multiple rounds.
However, unlike centralized DAS,
each device decides whether or not it uploads
its measurement by itself in distributed DAS.
Thus, distributed DAS can be used
if the BS does not have any knowledge of measurements
(e.g., second order statistics of $\bx$ in
Gaussian DAS \cite{Choi19_IoT} or 
measurement vectors in DAS for sparse signals
\cite{Choi19}).
In distributed DAS, since the BS does not decide uploading orders,
random access can be used 
with the access probability of each node
that is decided by its local measurement.
In this section, we show how random access can be
integrated with DAS using multichannel ALOHA 
\cite{Shen03} \cite{Chang15}.

%To help decide, the BS can feed back
%a global variable in each iteration.

Suppose that the BS wants to find $\by$ that is given by
\be
\by = G(\bx),
\ee
where $G(\cdot)$ is a certain function.
For example, a linear combination can be considered
as follows:
\be
\by = \sum_{k=1}^K \bg_k x_k = \bG \bx.
\ee
Let
$$
||\bg_{q(1)} x_{q(1)}||^2
\ge \ldots
\ge ||\bg_{q(K)} x_{q(K)}||^2,
$$
where $q(k)$ represents the index of the node
that has the $k$th largest norm.
Then, the uploading order can be decided according to
$\{q(k)\}$.
Assume that the BS knows $\{\bg_k\}$, but not $\{x_k\}$
that are available at nodes.
If the BS knows the second order statistics of the $x_k$'s,
$||\bg_k||^2 \uE[|x_k|^2]$ can replace
$||\bg_k||^2 |x_k|^2$ and the uploading order for DAS
can be decided by the BS.
However,
the BS may not have any information
of the $x_k$'s, and, furthermore, there might be
some nodes without any meaningful measurements
(i.e., $|x_k| \le \epsilon$).
Thus, it would be desirable to decide
the uploading order by the nodes (i.e., distributed DAS
is desirable).
In distributed DAS,
since the BS does not decide which devices are to transmit,
\emph{random access} can be employed for uploading,
where the access probability can be decided by each device
(based on its local measurement).

Suppose that there are
$L$ parallel multiple access channels for multichannel ALOHA.
Each user can randomly choose one of $L$ channels
if it decides to upload its measurement.
Let $p_k$ denote the access probability
of node $k$.
Let $q_k$ be the probability that node $k$
can successfully transmit its measurement without collision.
If there are multiple users that choose the same channel,
there is collision and no user succeeds to upload its measurement.
Then, we have
\be
q_k = p_k
\prod_{l \ne k} \left(1 - \frac{p_l}{L} \right).
\ee
In order to decide $p_k$ for given local measurement $x_k$
at node $k$, consider $\by$ 
after iteration $t$ as follows:
$$
\by = \by_1 + \by_2,
$$
where $\by_1 = \bG_1 \bx_1$ with $\bx_1$ corresponding to $\cX(t)$
and $\by_2 = \bG_2 \bx_2$ with $\bx_2$ corresponding to $\cX^c(t)$.
While $\by_1$
is known at the BS, we now
consider an approximation of $\by_2$ with $\bu$ from new 
uploaded measurements, which is given by
\begin{align}
\bu = \sum_{k \in \cK^c (t)} \bw_k \delta_k,
\end{align}
where $\bw_k = \bg_k x_k$, and $\delta_k \in \{0,1\}$ becomes
1 if the BS receives $x_k$ successfully and 0
otherwise.
The average number of successful uploading
with $L$ channels become
\begin{align}
\sum_k q_k
& = \sum_p p_k \prod_{l \ne k} \left(1 - \frac{p_l}{L} \right) \cr
& \le \sum_p p_k \exp \left(-\sum_{l \ne k}\frac{p_l}{L} \right)
\approx \sum_p p_k e^{-\frac{P}{L}} \cr
& = P e^{-\frac{P}{L}},
\end{align}
where $P = \sum_k p_k$.
Thus, $P$ is to be $L$ for maximizing the average number
of successful uploading, i.e.,
\be
\sum_{k} p_k = L \ \mbox{and} \ \sum_k q_k \le L e^{-1}.
\ee
Using the triangle inequality,
the expectation of the error norm
is bounded as
\begin{align}
\uE[||\by_2- \bu ||]
& = \uE[ ||\sum_{k \in \cK^c (t)} \bw_k (1 - \delta_k) || ] \cr
& \le \sum_{k \in \cK^c (t)} ||\bw_k|| \uE[ 1 - \delta_k] ] \cr
& = \sum_{k \in \cK^c (t)} ||\bw_k|| (1 - q_k)  \cr
& \le \sum_{k \in \cK^c (t)} ||\bw_k|| e^{- q_k},
	\label{EQ:bb}
\end{align}
where the last inequality is due to 
$1 - x < e^{-x}$, $x \ge 0$.
From \eqref{EQ:bb},
we can have the following 
optimization problem:
\begin{eqnarray}
& \min_{\{q_k\}} \sum_{k \in \cK^c (t)} ||\bw_k|| e^{-q_k} & \cr
& \mbox{subject to} \
\sum_{k \in \cK^c (t)} q_k \le L e^{-1}, \ q_k \in [0,e^{-1}]. &
\end{eqnarray}
When $\sum_k p_k = L$ to maximize
the average number of uploading,
it can be shown that
\be
q_k = p_k \prod_{l \ne k}
\left(1 - \frac{p_l}{L} \right) \approx p_k e^{-
\frac{ \sum_{l \ne k} p_l}{L}} \approx p_k e^{-1}.
\ee
Then, the optimization problem can be modified as
\begin{eqnarray}
& \min_{\{p_k\}} \sum_{k \in \cK^c (t)} ||\bw_k||
\exp(-e^{-1} p_k) & \cr
& \mbox{subject to} \
\sum_{k \in \cK^c (t)} p_k \le L, \ p_k \in [0,1]. &
\end{eqnarray}

It can be shown that
\be
p_k = \left[ e \ln ||\bw_k|| - \psi \right]_0^1,
  	\label{EQ:est}
\ee
where $\psi$ is a Lagrange multiplier, which can be decided
as
\be
\psi_{t+1} = \psi_t + \mu (\hat P_t - L),
\ee
based on dual ascent \cite{Boyd11}.
Here, $\hat P_t$ is the number of the nodes
that transmit measurements (regardless of collision),
which is available at the BS after each iteration.
The BS sends $\psi$
through feedback channels to the devices. Devices
find their access probability as in
\eqref{EQ:est} and upload their measurements accordingly.

For simulations, consider Bernoulli-Gaussian for $x_k$:
$$
x_k = \left\{
\begin{array}{ll}
\cN (0, 1), & \mbox{w.p. $p_{\rm s}$}\cr
0, & \mbox{w.p. $1 - p_{\rm s}$,}\cr
\end{array}
\right.
$$
where $p_{\rm s}$ denotes the probability of significant
measurement (i.e., $x_k$ has a large amplitude
with probability $p_{\rm s}$).
In this case, with probability
$1 - p_{\rm s}$, a node has a negligible measurement.
In addition, let
$$
[\bG]_{m,k} \sim \cN(0,1).
$$
Two different
distributed DAS approaches are considered as follows:
\begin{itemize}
\item Random Access 1 with measurement
independent access probability, $p_k = \frac{L}{|\cK^c(t)|}$
\item Random Access 2 with measurement
dependent access probability, $p_k =
\left[ e \ln ||\bw_k|| - \psi \right]_0^1$
\end{itemize}
Certainly, we expect that 
Random Access 2
can perform better than Random Access 1 if
the access probability is optimized.

\begin{figure}[thb]
\begin{center}
\includegraphics[width=\figwidth]{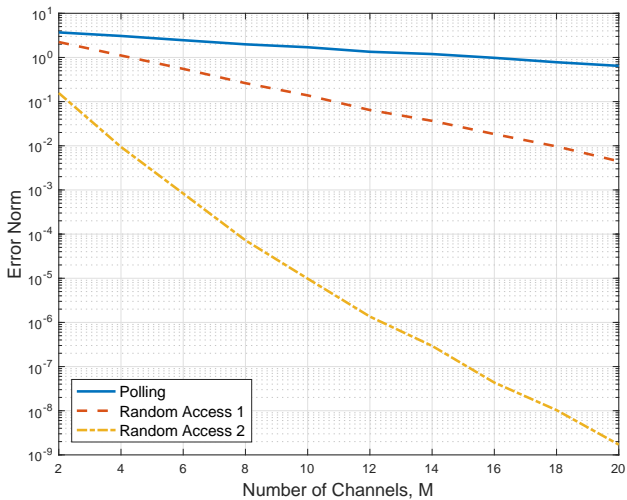}
\end{center}
\caption{Estimation Errors as functions of iterations
with $K = 400$, $L = 10$, $p_{\rm s} = 0.25$, and $\mu = 0.1$.}
        \label{Fig:plt1}
\end{figure}

In Fig.~\ref{Fig:plt1},
the estimation errors, $||\by - \by(t)||$,
are shown as functions of iterations, $t$,
with $K = 400$, $L = 10$, $p_{\rm s} = 0.25$, and $\mu = 0.1$.
As expected, 
Random Access 2, which has optimized access probabilities according
to their local measurements,
outperforms Random Access 1.

In \cite{Choi19_WCL},
we apply distributed DAS to federated learning
where devices upload their local
updates through parallel multiple access
channels. It is shown that Random Access 2
results in fast federated learning.

\section{Concluding Remarks}

In this paper, we presented the key idea of
DAS where sensing and communication meet for efficient
uploading from a large number of devices for
certain real-time IoT applications including federated learning.
We also discussed two
different types of DAS: centralized
and distributed DAS. In centralized DAS,
it was assumed that a BS has certain knowledge
of local measurements
(e.g., second order statistics of devices' measurements)
to decide uploading order. On the other hand,
in distributed DAS, the BS does not need to have
any knowledge of local measurements.

In distributed DAS, multichannel random
access was considered, since the BS does not decide
uploading order. By forming an optimization problem,
it was able to decide the access probability at each device
for efficient uploading with a limited bandwidth.

While we mainly focused on introducing  DAS
in terms of sensing and communication in this paper,
we did not discuss its practical applications in details.
We believe that DAS can be used for a number of different applications.
For example, for an application that provides user's journey time,
it may require to know traffic conditions 
that can be estimated from real-time updating of users' locations
(or vehicle's locations).
Since it may not be possible to collect the data set of
all users' locations in real-time (due to bandwidth limitation,
a large number of users, communication constraints, and so on),
DAS can be employed to 
predict the journey time of a requested trip
from a small number of 
\emph{selected} users' locations
that affect the journey time.
There would be a number of applications where DAS can 
efficiently
provide answers to real-time queries with sensors' or devices'
local data sets.

\bibliographystyle{ieeetr}
\bibliography{sensor}

\end{document}